\pdfoutput=1

\documentclass[10pt,twocolumn,twoside]{IEEEtran}

\usepackage{cite}

\usepackage[pdftex]{graphicx}
\usepackage[caption=false,font=footnotesize]{subfig}

\usepackage[cmex10]{amsmath}
\usepackage{amssymb}

\usepackage[noend]{algorithmic}

\usepackage{array}
\usepackage{multirow}

\usepackage{fixltx2e}

\usepackage{url}
\usepackage{hyperref}

\DeclareMathOperator{\E}{E}
\DeclareMathOperator{\tr}{tr}
\DeclareMathOperator{\Normal}{\mathcal{N}}

\newcommand{\R}{\mathbb{R}}
\newcommand{\given}{\,|\,}
\renewcommand{\vec}[1]{\boldsymbol{\mathbf{#1}}}

\begin{document}

\title{Dynamic stochastic blockmodels for \\time-evolving social networks}

\author{Kevin~S.~Xu,~\IEEEmembership{Member,~IEEE,}
        and~Alfred~O.~Hero~III,~\IEEEmembership{Fellow,~IEEE}%
\thanks{This work was partially supported by the 
Army Research Office grant W911NF-12-1-0443. 
K.~S.~Xu was partially supported by an award from the Natural Sciences and 
Engineering Research Council of Canada. 
Preliminary parts of this work appeared in the conference 
publication \cite{Xu2013a}.}%
\thanks{K.~S.~Xu is with the Technicolor Palo Alto Research Center, 
735 Emerson St, Palo Alto, CA 94301, USA 
(e-mail: \url{kevinxu@outlook.com})}%
\thanks{A.~O.~Hero III is with the Department of Electrical Engineering and 
Computer Science, University of Michigan, 1301 Beal Avenue, Ann Arbor, 
MI 48109-2122, USA 
(e-mail: \url{hero@umich.edu})}}

\maketitle

\begin{abstract}
Significant efforts have gone into the development of statistical models for 
analyzing data in the form of networks, such as social networks. 
Most existing work has focused on modeling static networks, which represent 
either a single time snapshot or an aggregate view over time. 
There has been recent interest in statistical modeling of 
\emph{dynamic networks}, 
which are observed at multiple points in time and offer a richer 
representation of many complex phenomena. 
In this paper, we present a state-space model for dynamic networks that 
extends the well-known \emph{stochastic blockmodel} for static networks 
to the dynamic setting. 
We fit the model in a near-optimal manner using an 
extended Kalman filter (EKF) augmented with a local search. 
We demonstrate that the EKF-based algorithm performs competitively with a 
state-of-the-art algorithm based on Markov chain Monte Carlo sampling 
but is significantly less computationally demanding.
\end{abstract}

\begin{IEEEkeywords}
State-space social network model, dynamic network, on-line estimation, 
extended Kalman filter
\end{IEEEkeywords}

\section{Introduction}
The study of networks has emerged as a topic of great interest in recent 
years. 
Many complex physical, biological, and social phenomena ranging from 
protein-protein interactions to the formation of social acquaintances can be 
naturally represented by networks. 
Analysis of data in the form of networks has recently captured the attention 
of the signal processing community 
\cite{Miller2010,Mutlu2012,Zhu2012a,Sandryhaila2013,Park2013}. 
To date, much research has focused on static networks, which 
either represent a single time snapshot of the phenomenon  
of interest or an aggregate view over time. 
Accordingly, many statistical models for networks have been developed for 
static networks; see
\cite{Goldenberg2009} for a survey of the literature. 
However, most complex phenomena, including social behavior, 
are time-varying, which has led researchers in recent years to examine 
\emph{dynamic, time-evolving networks}. 
Previous studies have typically examined aspects of dynamic networks 
related to 
their growth over time, including densification \cite{Katz2005,Leskovec2007} 
and shrinking diameters \cite{Leskovec2007}, and to their structural changes, 
including the temporal evolution of communities in the network 
\cite{Mucha2010,Greene2010,Xu2013}. 

In this paper, we consider dynamic networks represented 
by a sequence of snapshots of the network at discrete time steps. 
Both nodes and edges of the network could be \emph{added or removed} 
over time. 
In a dynamic social network, for example, the nodes could correspond to 
people, and the edges could correspond to interactions between people. 
The presence of an edge between two nodes $i$ and $j$ at time 
step $t$ would then indicate that an interaction between $i$ and $j$ occurred 
during the time window represented by time step $t$. 

We characterize such dynamic networks using a set of unobserved 
\emph{time-varying 
states} from which the observed snapshots are derived. 
We utilize a state-space model for dynamic networks first proposed in 
\cite{Xu2013a} that combines two types 
of statistical models: a static model for the individual 
snapshots and a temporal model for the evolution of the states. 
The network snapshots are modeled using the \emph{stochastic blockmodel} (SBM) 
\cite{Holland1983}, a simple parametric model commonly used in the analysis 
of static social networks. 
The state evolution is modeled by a stochastic dynamic system. 

Using a Gaussian approximation, which becomes increasingly accurate as the 
SBM block sizes increase, we employ a \emph{near-optimal} 
procedure for fitting the proposed model in the 
\emph{on-line} setting where only 
past and present network snapshots are available. 
The inference procedure consists of an extended Kalman 
filter (EKF) \cite{Haykin2001} augmented with a local search strategy. 
The proposed algorithm is considerably faster than a state-of-the-art 
algorithm that uses Markov chain Monte Carlo (MCMC) sampling, yet  
our experiments show that the proposed algorithm has
comparable accuracy to the more computationally demanding MCMC-based algorithm 
in recovering the true states. 
We apply the proposed algorithm to analyze the Enron network 
\cite{Priebe2005,Priebe2009},  
a dynamic social network of email communication, and reveal several 
interesting trends that cannot be identified by aggregate statistics such 
as the total number of emails sent at each time step. 

We first proposed the main elements of our model, including the Gaussian 
approximation and EKF algorithm for tracking dynamic SBMs, in \cite{Xu2013a}. 
This paper refines and extends the analysis of the model in several ways. 
First, a detailed development of the approximate inference 
procedure is given, including a study of the approximation accuracy. 
Second, a more extensive performance analysis is presented, for both simulated 
and real data, that establishes the advantages of the proposed dynamic SBM 
method relative to other methods.

\section{Background}
\label{sec:Background}

\subsection{Notation}
Time steps are denoted by superscripts, while matrix or vector indices are 
denoted by subscripts. 
We represent a dynamic network by a time-indexed sequence of graphs, with 
$W^t = [w_{ij}^t]$ denoting the adjacency matrix of the graph 
observed at time step $t$. 
We define 
$w_{ij}^t = 1$ if there is an edge from node $i$ to node $j$ at time $t$, 
and $w_{ij}^t = 0$ otherwise. 
We assume that the graphs are directed, 
i.e.~$w_{ij}^t \neq w_{ji}^t$ in general, and that 
there are no self-edges, i.e.~$w_{ii}^t = 0$. 
$W^{(t)}$ denotes the set of all snapshots up to time $t$, $\{W^t, 
W^{t-1}, \ldots, W^1\}$. 
$|V^t|$ and $|E^t|$ denote the number of observed nodes and edges, 
respectively, at time $t$. 
We write $X^{t_1|t_2}, t_1 \geq t_2$ to denote a quantity $X$ at time $t_1$ 
computed using only observations from time $t_2$ and earlier. 
The notation $i \in a$ indicates that node $i$ is a member of class $a$. 
$|a|$ denotes the number of nodes in class  $a$. 
The classes of all nodes at time $t$ is given by a vector 
$\vec{c}^t$ with $c_i^t = a$ if $i \in a$ at time $t$. 
We denote the submatrix of $W^t$ corresponding to the relations between 
nodes in class $a$ and nodes in class $b$ by $W_{[a][b]}^t$. 
Finally, we denote the vectorized equivalent of a matrix $X$, i.e.~the vector 
obtained by simply stacking columns of $X$ on top of one another, by 
$\vec{x}$. 
Doubly-indexed subscripts such as $x_{ij}$ denote entries of matrix $X$, 
while singly-indexed subscripts such as $x_i$ denote entries of the 
vectorized equivalent $\vec{x}$. 

\subsection{Static stochastic blockmodels}
\label{sec:Static_SBM}
We present a brief summary of the \emph{static 
stochastic blockmodel} (SSBM) \cite{Holland1983}, which we use 
as the static model for individual network snapshots. 
Consider a snapshot at an arbitrary time step $t$. 
An SSBM is parameterized by a $k \times k$ matrix $\Theta^t = [\theta_{ab}^t]$, 
where $\theta_{ab}^t$ denotes the \emph{probability of forming an edge} 
between a node in 
class $a$ and a node in class $b$, and $k$ denotes the number of classes. 
The SSBM decomposes the adjacency matrix into $k^2$ blocks, where each block 
is associated with relations between nodes in classes $a$ and $b$. 
Each block $(a,b)$ corresponds to a submatrix $W_{[a][b]}^t$ of the adjacency 
matrix $W^t$.
Thus, given the \emph{class membership vector} $\vec{c}^t$, 
each entry of $W^t$ is 
an independent realization of a Bernoulli random variable with a 
block-dependent parameter; that is, $w_{ij}^t \sim \text{Bernoulli} 
\Big(\theta_{c_i^t c_j^t}^t\Big)$. 

SSBMs are used in two settings:
the \emph{a priori} blockmodeling setting, 
where class memberships are known or assumed, and the objective is to 
estimate the matrix of edge probabilities $\Theta^t$, and 
the \emph{a posteriori} blockmodeling setting, where the objective 
is to simultaneously estimate $\Theta^t$ and the class membership 
vector $\vec{c}^t$. 
Since each entry of $W^t$ is independent, the likelihood for the 
parameters $\Phi^t$ of the SSBM is given by
\begin{equation}
	\label{eq:Likelihood_Phi}
	f\left(W^t; \Phi^t\right) = \prod_{i \neq j} \left( 
		\theta_{c_i^t c_j^t}^t\right)^{w_{ij}^t} \left(1 - 
		\theta_{c_i^t c_j^t}^t\right)^{1-w_{ij}^t}.
\end{equation}
The likelihood \eqref{eq:Likelihood_Phi} can be rewritten as
\begin{equation}
	\label{eq:Likelihood_Phi_simp}
	\begin{split}
	f\left(W^t; \Phi^t\right) = \exp\Bigg\{&\sum_{a=1}^k \sum_{b=1}^k
		\big[m_{ab}^t \log \left(\theta_{ab}^t\right) \\
	&+ \left(n_{ab}^t - 
		m_{ab}^t\right) \log \left(1 - \theta_{ab}^t\right)\big]\Bigg\},
	\end{split}
\end{equation}
where $m_{ab}^t = \sum_{i \in a} \sum_{j \in b} w_{ij}^t$ denotes the number 
of \emph{observed edges} in block $(a,b)$, and 
\begin{equation}
	\label{eq:N_def}
	n_{ab}^t = 
	\begin{cases}
		|a||b| & a \neq b \\
		|a|(|a|-1) & a = b
	\end{cases}
\end{equation}
denotes the number of \emph{possible edges} in block $(a,b)$ 
\cite{Karrer2011}.
The parameters are given by $\Phi^t = \Theta^t$ in the a priori setting, and 
$\Phi^t = \{\Theta^t,\vec{c}^t\}$ in the a posteriori setting. 
In the a priori setting, a sufficient statistic for estimating $\Theta^t$ is 
the matrix $Y^t$ of \emph{block densities} corresponding to ratios of observed 
edges relative to possible edges within each block, which has entries 
$y_{ab}^t = m_{ab}^t / n_{ab}^t$.
The matrix $Y^t$ is also the maximum-likelihood estimate of $\Theta^t$ 
\cite{Karrer2011}. 

Parameter estimation in the a posteriori setting is more involved, and 
many methods have been proposed, including Gibbs sampling 
\cite{Nowicki2001}, 
label-switching \cite{Karrer2011,Zhao2012}, and spectral clustering 
\cite{Rohe2011,Sussman2012}. 
The label-switching methods use a heuristic for solving the combinatorial 
optimization problem of maximizing the likelihood 
\eqref{eq:Likelihood_Phi_simp} 
over the set of possible class memberships, which is too large 
for an exhaustive search to be tractable. 
The spectral clustering methods utilize the eigenvectors of the adjacency 
matrix $W^t$ or a similar matrix to estimate the class memberships. 

\subsection{Related work}
\label{sec:Related_work}

Several statistical models for dynamic networks have previously been 
proposed for modeling and tracking dynamic networks \cite{Goldenberg2009}. 
Guo et al.~\cite{Guo2007} 
proposed a temporal extension of the exponential random 
graph model (ERGM) called the hidden temporal ERGM. 
Sarkar and Moore \cite{Sarkar2005} 
proposed a temporal extension of the latent space network model 
and developed an algorithm to compute point estimates of node 
positions over time using 
conjugate gradient optimization initialized from a 
multidimensional scaling solution. 
In \cite{Sarkar2007}, Sarkar et al.~proposed a Gaussian approximation that 
allowed for approximate inference on the dynamic latent space model using 
Kalman filtering. 
The approach of \cite{Sarkar2007} is similar in flavor to the approach 
we employ in this paper; however, our approach involves a different static 
model, namely the stochastic blockmodel, for the network snapshots 
and uses this model to develop an extended Kalman filter (EKF) to track 
the model parameters. 

Hoff \cite{Hoff2011} proposed a dynamic latent factor model analogous to an 
eigenvalue decomposition with time-invariant eigenvectors and time-varying 
eigenvalues. 
The model is applicable to many types of data in the form of multi-way 
arrays, including dynamic social networks, and is fit using MCMC sampling. 
In \cite{Lee2011}, Lee and Priebe proposed a latent process model for 
attributed 
(multi-relational) dynamic networks using random dot product spaces. 
The authors fit mathematically tractable 
first- and second-order approximations of the 
random dot process model, for which individual network snapshots are 
drawn from attributed versions of the Erd\H{o}s-R\'{e}nyi and 
latent space models, respectively. 
Perry and Wolfe \cite{Perry2013} proposed a point process model for 
dynamic networks of directed interactions and a partial likelihood inference 
procedure to fit their model. 
The authors model interactions using a multivariate 
counting process that accounts for effects including homophily. 
Their model operates in continuous time, unlike the proposed model in this 
paper, which operates on discrete-time snapshots. 

More closely related to the state-space dynamic network model 
we consider in this paper are several temporal 
extensions of stochastic blockmodels (SBMs). 
Xing et al.~\cite{Xing2010} and Ho et al.~\cite{Ho2011} proposed 
temporal extensions of a mixed-membership version of the SBM 
using linear state-space models for the real-valued class memberships. 
In \cite{Yang2011}, Yang et al.~proposed a temporal extension of the SBM 
that is  similar to our proposed model. 
The main difference is that the authors explicitly modeled nodes changing 
between classes over time by using a transition matrix that 
specifies the probability that a node in class $i$ at time $t$ switches to 
class $j$ at time $t+1$ for all $i,j,t$. 
The authors fit the model using a combination of 
Gibbs sampling and simulated annealing, which they refer to as 
\emph{probabilistic simulated annealing} (PSA). 
We use the performance of the PSA 
algorithm as a baseline for comparison with the 
less computationally demanding EKF-based approximate inference 
procedure we utilize in this paper. 

\section{Dynamic stochastic blockmodels}
\label{sec:Dynamic_SBM}

We now present a state-space model for dynamic networks that accomplishes 
a temporal extension of the SSBM. 
First we review the dynamic SBM and approximate inference procedure 
for both a priori and a posteriori blockmodeling proposed in \cite{Xu2013a}. 
Next we present an analysis of the time complexity of the inference 
procedure and discuss hyperparameter estimation. 
Finally we investigate the validity and accuracy of two key approximations 
used in the inference procedure. 

\subsection{A priori blockmodels}
\label{sec:Inference_known}
In the a priori SSBM setting, $Y^t$ is a sufficient statistic for 
estimating $\Theta^t$ as discussed in Section \ref{sec:Static_SBM}. 
The entries of $W_{[a][b]}^t$ are independent and identically distributed 
(iid) $\text{Bernoulli}\left(\theta_{ab}^t\right)$. 
Thus the sample mean $y_{ab}^t$ follows a re-scaled binomial distribution. 
For large block size, $y_{ab}^t$ is approximately 
Gaussian by the Central Limit Theorem with mean $\theta_{ab}^t$ and variance 
\begin{equation}
	\label{eq:Sigma_t_Theta}
	(\sigma_{ab}^t)^2 = \theta_{ab}^t(1-\theta_{ab}^t) / n_{ab}^t,
\end{equation}
where $n_{ab}^t$ was defined in \eqref{eq:N_def}. 
We assume that $y_{ab}^t$ is indeed Gaussian for all $(a,b)$ and posit 
the linear observation model $Y^t = \Theta^t + Z^t$, 
where $Z^t$ is a zero-mean independent 
Gaussian noise matrix with variance $(\sigma_{ab}^t)^2$ for the 
$(a,b)$th entry. 

In the dynamic setting where past snapshots are 
available, the observations would be given by the set $Y^{(t)}$. 
The set $\Theta^{(t)}$ can then be viewed as the states of a dynamic 
system that generates the noisy observation sequence. 
We complete the model by specifying a model for the state evolution over time. 
Since $\theta_{ab}^t$ is a probability and must be bounded between $0$ and $1$, 
we instead work with the matrix $\Psi^t = [\psi_{ab}^t] \in \R^{k \times k}$ 
where $\psi_{ab}^t = 
\log(\theta_{ab}^t) - \log(1 - \theta_{ab}^t)$, the logit of $\theta_{ab}^t$. 
A simple model for the state evolution is given by the linear dynamic system 
\begin{equation}
	\label{eq:State_evol_model}
	\vec{\psi}^t = F^t \vec{\psi}^{t-1} + \vec{v}^t,
\end{equation}
where $F^t$ is the state transition model applied to the previous state, 
$\vec{\psi}^t$ is the vector representation of the matrix $\Psi^t$, and 
$\vec{v}^t$ is a random vector of zero-mean Gaussian entries, 
commonly referred to as process noise, with covariance matrix $\Gamma^t$. 
The entries of the process noise vector $\vec{v}^t$ are not assumed to be 
independent (unlike the entries of $Z^t$, which are independent 
by construction) to allow for states to evolve in a correlated manner. 
The state transition matrix $F^t$ may either be known, as in the case 
of structural time series models \cite{Durbin2012}, or can be estimated 
using methods for system identification \cite{Ljung1999}. 
For the simplest model where $F^t = I$ for all $t$, the state evolution 
follows a multivariate Gaussian random walk. 

\begin{figure}[t]
	\centering
	\includegraphics[width=3.4in]{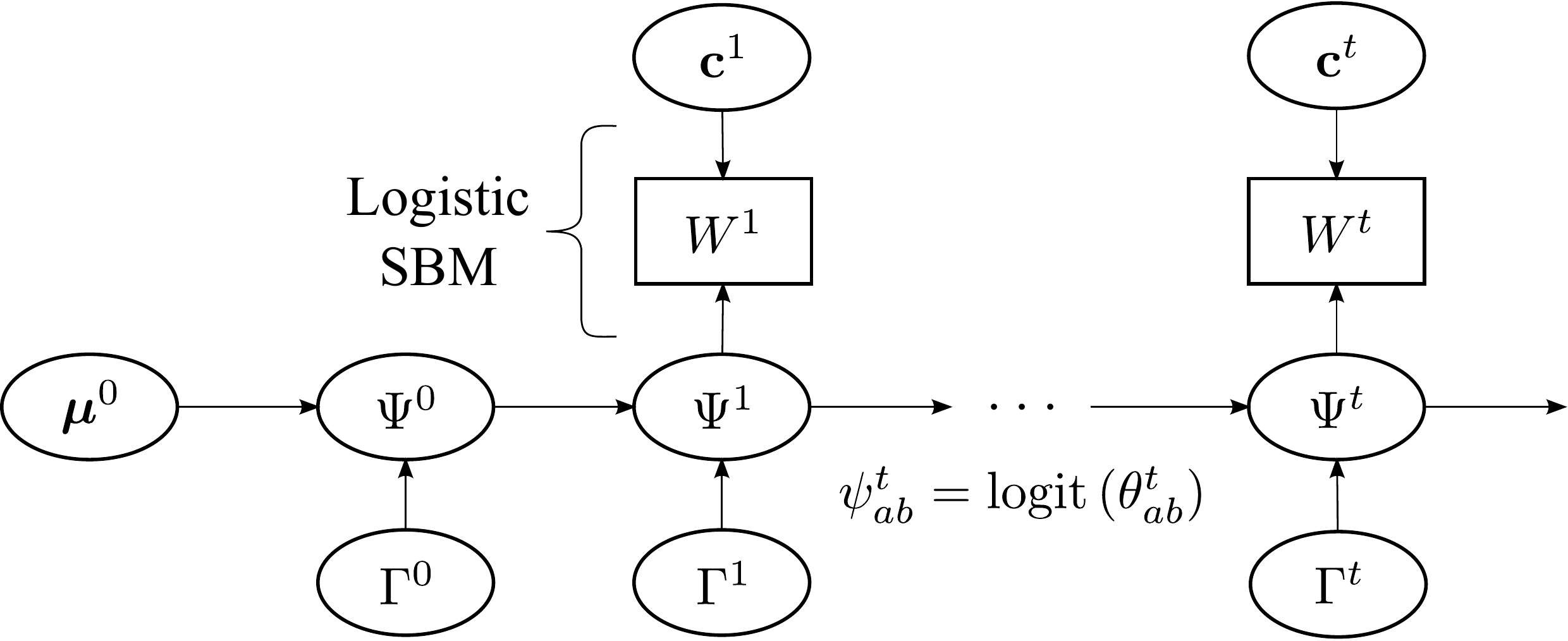}
	\caption{Graphical representation of the dynamic SBM. 
		The rectangular boxes denote observed quantities, and the ovals 
		denote unobserved quantities.
		The logistic SBM refers to applying the logistic function 
		\eqref{eq:Logistic_fn} to each 
		entry of $\Psi^t$ to obtain $\Theta^t$ then generating $W^t$ using 
		$\Theta^t$ and $\vec{c}^t$.}
	\label{fig:Graphical_model_proposed}
\end{figure}

The observation model can be written 
in terms of $\vec{\psi}^t$ as\footnote{Note that we have converted the block 
densities $Y^t$ and observation noise $Z^t$ to their respective vector 
representations $\vec{y}^t$ and $\vec{z}^t$.}
\begin{equation}
	\label{eq:Obs_model_SBM}
	\vec{y}^t = h\left(\vec{\psi}^t\right) + \vec{z}^t,
\end{equation}
where the function $h: \R^{k^2} \rightarrow \R^{k^2}$ is defined by 
\begin{equation}
	\label{eq:Logistic_fn}
	h_i(\vec{x}) = 1/(1+e^{-x_i}), 
\end{equation}
i.e.~the logistic function applied to each entry of $\vec{x}$. 
We denote the covariance matrix of $\vec{z}^t$ by $\Sigma^t$, which is a 
diagonal 
matrix\footnote{The indices $(a,b)$ 
for $(\sigma_{ab}^t)^2$ are converted into a single index $i$ 
corresponding to the vector representation $\vec{z}^t$.} with entries given by 
$(\sigma_{ab}^t)^2$. 
A graphical representation of the dynamic network model 
is shown in Fig.~\ref{fig:Graphical_model_proposed}. 

The inference procedure for the model is \emph{on-line}, 
i.e.~the state estimate at time $t$ is 
formed using only observations from time $t$ and earlier. 
We assume the initial state is Gaussian distributed, 
i.e.~$\vec{\psi}^0 \sim \Normal\left(\vec{\mu}^0, \Gamma^0\right)$, and that 
$\{\vec{\psi}^0, \vec{v}^1, 
\ldots, \vec{v}^t, \vec{z}^1, \ldots, \vec{z}^t\}$ are mutually independent. 
If \eqref{eq:Obs_model_SBM} was linear in $\vec{\psi}^t$, then the optimal 
estimate of $\vec{\psi}^t$ in terms of minimizing mean-squared error 
would be given by the Kalman filter \cite{Haykin2001}. 
Due to the non-linearity, we apply the extended Kalman filter (EKF), which 
linearizes the dynamics about the predicted state and provides a 
\emph{near-optimal} estimate of $\vec{\psi}^t$. 
The EKF equations for the model specified by \eqref{eq:State_evol_model}, 
\eqref{eq:Obs_model_SBM} are as follows. 
The predicted state estimate is 
\begin{equation}
	\label{eq:EKF_prior_state}
	\hat{\vec{\psi}}^{t|t-1} = F^t \hat{\vec{\psi}}^{t-1|t-1},
\end{equation}
and the predicted covariance estimate is
\begin{equation}
	\label{eq:EKF_prior_cov}
	R^{t|t-1} = F^t R^{t-1|t-1} \left(F^t\right)^T + \Gamma^t. 
\end{equation}
Let $H^t$ denote the Jacobian of $h$ evaluated at the predicted state 
estimate $\hat{\vec{\psi}}^{t|t-1}$. 
$H^t$ is a diagonal matrix with $(i,i)$th entry given by 
\begin{equation}
	\label{eq:Logistic_Jacobian}
	h'_i\left(\hat{\vec{\psi}}_i^{t|t-1}\right) 
		= \frac{\exp\left(-\hat{\vec{\psi}}_i^{t|t-1}\right)} 
		{\left[1 + \exp\left(-\hat{\vec{\psi}}_i^{t|t-1}\right)\right]^2}.
\end{equation}
The near-optimal (when the estimation errors are small enough to make the 
EKF linearization accurate) Kalman gain is given by
\begin{equation}
	\label{eq:EKF_Kalman_gain}
	K^t = R^{t|t-1} \left(H^t\right)^T 
		\left[H^t R^{t|t-1} \left(H^t\right)^T + \Sigma^t\right]^{-1},
\end{equation}
from which the updated state estimate
\begin{equation}
	\label{eq:EKF_post_state}
	\hat{\vec{\psi}}^{t|t} 
		= \hat{\vec{\psi}}^{t|t-1} + K^t\left[\vec{y}^t 
		- h\left(\hat{\vec{\psi}}^{t|t-1}\right)\right]
\end{equation}
and the updated covariance estimate
\begin{equation}
	\label{eq:EKF_post_cov}
	R^{t|t} = \left(I - K^t H^t\right) R^{t|t-1}
\end{equation}
are obtained \cite{Haykin2001}.
The updated state estimate $\hat{\vec{\psi}}^{t|t}$ provides a near-optimal  
fit to the model at time $t$ given the observed sequence $W^{(t)}$. 

\subsection{A posteriori blockmodels}
\label{sec:Inference_unknown}

\begin{figure}[t]
	\begin{algorithmic}[1]
	\STATE $\hat{\vec{c}}^t \leftarrow \hat{\vec{c}}^{t-1}$ 
		\COMMENT{Initialize class memberships}
	\STATE Compute block densities $Y^t$ using $W^t$ and $\hat{\vec{c}}^t$
	\STATE Compute $\hat{\vec{\psi}}^{t|t}$ using EKF equations 
		\eqref{eq:EKF_prior_state}--\eqref{eq:EKF_post_cov}
	\STATE Compute log-posterior $p^t$ by substituting 
		$\hat{\vec{\psi}}^{t|t}$ for $\vec{\psi}^t$ in 
		\eqref{eq:Psi_posterior_simp}
	\WHILE[Local search algorithm]{$iter \leq max\_iter$}
	\STATE $\bar{p}^t \leftarrow -\infty$ \COMMENT{Log-posterior 
		of best neighboring solution up to a constant}
	\STATE $\tilde{\vec{c}}^t \leftarrow \hat{\vec{c}}^t$ \COMMENT{Solution 
		currently being visited}
	\FOR[Visit all neighboring solutions]{$i = 1$ \TO $|V^t|$}
		\FOR{$j = 1$ \TO $k$ such that $\hat{c}_i^t \neq j$}
			\STATE $\tilde{c}_i^t \leftarrow j$	\COMMENT{Change class of a 
				single node}
			\STATE Compute block densities $\tilde{Y}^t$ using $W^t$ and 
				$\tilde{\vec{c}}^t$
			\STATE Compute $\tilde{\vec{\psi}}^{t|t}$ using EKF equations 
				\eqref{eq:EKF_prior_state}--\eqref{eq:EKF_post_cov}
			\STATE Compute log-posterior $\tilde{p}^t$ by substituting 
				$\tilde{\vec{\psi}}^{t|t}$ for $\vec{\psi}^t$ in 
				\eqref{eq:Psi_posterior_simp}
			\IF[Visited solution is better than best neighboring solution] 
				{$\tilde{p}^t > \bar{p}^t$}
				\STATE $\big[\bar{p}^t,\bar{\vec{\psi}}^{t|t}, 
					\bar{\vec{c}}^t\big] \leftarrow \big[\tilde{p}^t, 
					\tilde{\vec{\psi}}^{t|t},\tilde{\vec{c}}^t\big]$
			\ENDIF
			\STATE $\tilde{c}_i^t \leftarrow \hat{c}_i^{t}$ \COMMENT{Reset 
				class membership of current node}
		\ENDFOR
	\ENDFOR
	\IF[Best neighboring solution is better than current best solution]  
		{$\bar{p}^t > p^t$}
		\STATE $\big[p^t,\hat{\vec{\psi}}^{t|t}, 
			\hat{\vec{c}}^t\big] \leftarrow \big[\bar{p}^t, 
			\bar{\vec{\psi}}^{t|t},\bar{\vec{c}}^t\big]$
	\ELSE[Reached local maximum]
		\STATE \textbf{break}
	\ENDIF
	\ENDWHILE
	\RETURN $\big[\hat{\vec{\psi}}^{t|t},\hat{\vec{c}}^t\big]$
	\end{algorithmic}
	\caption{A posteriori blockmodel inference procedure at time $t$ 
		using the EKF.}
	\label{alg:EKF_post_alg}
\end{figure}

In many applications, the class memberships $\vec{c}^t$ are not known a 
priori and must be estimated along with $\Psi^t$. 
This can be done using label-switching methods as in 
\cite{Karrer2011,Zhao2012}, but rather than maximizing 
the likelihood \eqref{eq:Likelihood_Phi_simp}, we 
maximize the posterior state density given the entire sequence of 
observations $W^{(t)}$ up to time $t$ to account for the prior information. 
This is done by alternating between label-switching and applying the EKF 
to arrive at a maximum a posteriori probability (MAP) estimate of $\vec{c}^t$.

The posterior state density is given by
\begin{equation}
	\label{eq:Psi_posterior}
	f\big(\vec{\psi}^t \given W^{(t)}\big) \propto f\big(W^t 
		\given \vec{\psi}^t, W^{(t-1)}\big) f\big(\vec{\psi}^t \given 
		W^{(t-1)}\big).
\end{equation}
By the conditional independence of current and past observations given the 
current state, $W^{(t-1)}$ drops out of the first multiplicative factor on 
the right side of \eqref{eq:Psi_posterior}. 
This factor can thus be obtained simply by substituting $h(\Psi^t)$ 
for $\Theta^t$ in \eqref{eq:Likelihood_Phi_simp}. 
We approximate the second term in \eqref{eq:Psi_posterior} with 
$f\left(\vec{\psi}^t \given \vec{y}^{(t-1)}\right)$ using the estimated class 
memberships at all previous time steps. 
By applying the Kalman filter to the linearized temporal model 
\cite{Haykin2001}, $f\left(\vec{\psi}^t \given \vec{y}^{(t-1)}\right) \sim 
\Normal\big(\hat{\vec{\psi}}^{t|t-1}, R^{t|t-1}\big)$. 
Thus the logarithm of the posterior density is given by
\begin{equation}
	\label{eq:Psi_posterior_simp}
	p^t = \log f\big(W^t \given \vec{\psi}^t\big) 
		+ \log f\big(\vec{\psi}^t \given \vec{y}^{(t-1)}\big),
\end{equation}
where
\begin{equation*}
\begin{split}
	\log f\big(W^t \given \vec{\psi}^t\big) = \sum_{a=1}^k &\sum_{b=1}^k 
		\Big\{m_{ab}^t \log \left[h\left(\psi_{ab}^t\right)\right] \\
	&+ \left(n_{ab}^t - m_{ab}^t\right) 
		\log \left[1 - h\left(\psi_{ab}^t\right)\right]\!\Big\},
\end{split}
\end{equation*}
\begin{equation*}
\begin{split}
	\log f\big(\vec{\psi}^t \given \vec{y}^{(t-1)}\big) 
		= -&\frac{1}{2} \left(\vec{\psi}^t - \hat{\vec{\psi}}^{t|t-1} 
		\right)^T \\
	&\left(R^{t|t-1}\right)^{-1} \left(\vec{\psi}^t 
		- \hat{\vec{\psi}}^{t|t-1}\right) + \mathcal{C},
\end{split}
\end{equation*}
and $\mathcal{C}$ is a constant term independent of $\vec{\psi}^t$ that 
can be ignored\footnote{At the 
initial time step, $\hat{\vec{\psi}}^{1|0} = \vec{\mu}^0$ and 
$R^{1|0} = F^1\Gamma^0\left(F^1\right)^T+\Gamma^1$.}. 

\begin{figure}[t]
	\begin{algorithmic}[1]
	\STATE Compute singular value decomposition of $W^1$
	\STATE $\tilde{\Sigma} \leftarrow$ diagonal matrix of $k$ largest 
		singular values
	\STATE $(\tilde{U}, \tilde{V}) \leftarrow$ left and right 
		singular vectors for $\tilde{\Sigma}$
	\STATE $\tilde{Z} \leftarrow \big[\tilde{U} \tilde{\Sigma}^{1/2}, 
		\tilde{V} \tilde{\Sigma}^{1/2}\big]$ \COMMENT{concatenate 
		scaled left and right singular vectors}
	\STATE $\hat{\vec{c}}^0 \leftarrow$ k-means clustering on rows of 
		$\tilde{Z}$
	\RETURN $\hat{\vec{c}}^0$
	\end{algorithmic}
	\caption{SSBM spectral clustering initialization to obtain initial 
		estimate $\hat{\vec{c}}^0$ of class memberships at $t=1$.}
	\label{alg:SSBM_init}
\end{figure}

We use the log-posterior \eqref{eq:Psi_posterior_simp} 
as the objective function for label-switching to obtain 
an a posteriori fit to the dynamic SBM. 
We find that a simple local search (hill climbing) algorithm 
\cite{Russell2003} initialized using the 
estimated class memberships at the previous time step suffices, because 
only a small fraction of nodes change classes between time steps in most 
applications. 
Pseudocode for the a posteriori inference procedure is shown in 
Fig.~\ref{alg:EKF_post_alg}. 
At the initial time step, we employ the spectral clustering algorithm of 
Sussman et al.~\cite{Sussman2012} for the SSBM to generate an initial estimate 
$\hat{\vec{c}}^0$ of the class memberships at $t=1$. 
The spectral clustering algorithm is given in Fig.~\ref{alg:SSBM_init}. 
Using the spectral clustering solution as the initialization prevents the 
local search from getting stuck in a poor local maximum at the initial 
time step. 

\subsection{Time complexity}
\label{sec:Complexity}

We begin with an analysis of the time complexity of the inference procedure 
for a priori blockmodeling at each time step. 
Calculating the matrix of block densities $Y^t$ involves summing over all 
edges present at time $t$, which has $O(|E^t|)$ time complexity. 
Application of the EKF requires only matrix-vector multiplications and 
a matrix inversion (to calculate the Kalman gain). 
The size of both the observation and state vectors in the EKF is 
$k^2$, so the time complexity of the EKF is dominated by the 
$O(k^6)$ complexity of the matrix inversion. 
Hence the overall time complexity at time $t$ is $O(|E^t|+k^6)$. 

A posteriori blockmodeling involves performing a local search at each time 
step in addition to applying the EKF. 
At each iteration of the local search, all $|V^t|(k-1)$ neighboring class 
assignments are visited. 
For each class assignment, we compute the EKF estimate 
$\hat{\vec{\psi}}^{t|t}$ and substitute it into the log-posterior 
\eqref{eq:Psi_posterior_simp}. 
As previously mentioned, computing the EKF estimate is dominated by the 
$O(k^6)$ complexity of an inverting a $k^2 \times k^2$ matrix. 
Evaluating the log-posterior \eqref{eq:Psi_posterior_simp} 
also requires inversion of a 
$k^2 \times k^2$ matrix. 
The matrix inversions are independent of the class assignments so they 
only need to be performed once at each time step rather than at each 
iteration of the local search. 
Thus the time complexity at each local search iteration is reduced to $O(k^4)$ 
for the matrix-vector multiplications. 
The overall time complexity at time $t$ then becomes $O(|E^t|+k^6+|V^t|lk^5)$, 
where $l$ denotes the number of local search iterations. 
We note that the local search algorithm is easily parallelized; 
specifically, each visit to 
a neighboring class assignment can be executed on a separate core 
or processor, which 
can significantly reduce the run-time of a posteriori inference.

\subsection{Estimation of hyperparameters}
\label{sec:Hyperparameters}
The EKF-based inference procedure requires four hyperparameters 
to be specified:
\begin{enumerate}
	\item the mean $\vec{\mu}^0$ of the initial state $\vec{\psi}^0$,
	\item the covariance matrix $\Gamma^0$ of the initial state 
		 $\vec{\psi}^0$,
	\item the covariance matrix $\Sigma^t$ of the observation noise 
		$\vec{z}^t$, and 
	\item the covariance matrix $\Gamma^t$ of the process (state evolution) 
		noise $\vec{v}^t$.
\end{enumerate}

The first two hyperparameters relate to the initial state. 
In the absence of prior information about the network, specifically the 
matrix $\Theta^0$ of probabilities of forming edges, we employ a diffuse 
prior \cite{Durbin2012}; 
that is, we let the variances of the initial states approach $\infty$. 
This can be implemented by simply taking $\hat{\vec{\psi}}^{1|1} 
= g(\vec{y}^1)$ and and $R^{1|1} = G^1 \Sigma^1 (G^1)^T$, where 
$g_i(\vec{x}) = h_i^{-1}(x) = \log(x_i) - \log(1-x_i)$ is the logit of the 
$i$th entry of $\vec{x}$, 
and $G^1$ is the Jacobian of $g$ evaluated at $\vec{y}^1$, 
which is a diagonal matrix with 
entries given by $g_i'(\vec{y}^1) = 1/y_i^1 + 1/(1-y_i^1)$. 
Thus the initial state mean and covariance are given by the 
transformed initial observation and its covariance. 

The third hyperparameter $\Sigma^t$ denotes the covariance matrix of the 
observation noise. 
In many applications of state-space models, it is assumed to be 
time-invariant and estimated jointly with $\Gamma^t$. 
In the dynamic SBM setting, however, $\Sigma^t$ is necessarily 
time-varying because it is related to the current state 
$\vec{\psi}^t$ through \eqref{eq:Sigma_t_Theta} and the logistic function 
$h$ \eqref{eq:Logistic_fn}. 
Taking advantage of this relationship, we use a plug-in estimator for 
$\Sigma^t$ by substituting $h\big(\hat{\vec{\psi}}^{t|t-1}\big)$ for 
$\Theta^t$ in \eqref{eq:Sigma_t_Theta}. 

The final hyperparameter $\Gamma^t$ denotes the covariance matrix of the 
process noise $\vec{v}^t$. 
Unlike $\Sigma^t$, we assume $\Gamma^t$ to be time-invariant. 
Furthermore, it is not 
necessarily diagonal because states could evolve in a correlated manner. 
For example, if $\psi_{ab}^t$ increases from time $t-1$ to time $t$, it 
may be a result of some action by nodes in class $a$, which could also affect 
the other entries in row $a$ of $\Psi^t$. 
Although $\Gamma^t$ is not necessarily diagonal, it is desirable to impose 
some structure on $\Gamma^t$ so that one does not have to estimate all 
$O(k^4)$ covariances individually. 
Accordingly we assume the structure of $\Gamma^t$ is such that 
\begin{equation}
	\label{eq:Gamma_struct}
	\gamma_{ij}^t = 
	\begin{cases}
		s_{diag}, & i = j \\
		s_{nb}, & i,j \text{ are neighboring cells in } \Psi^t \\
		0, & \text{otherwise},
	\end{cases}
\end{equation}
where $i,j$ being neighboring cells means that the matrix indices $(a_i,b_i)$ 
corresponding to $i$ in $\Psi^t$ are in the same row or column as matrix 
indices $(a_j,b_j)$. 
This choice for $\Gamma^t$ exploits the fact that the state $\Psi^t$ is 
actually a matrix that has been flattened into a vector $\vec{\psi}^t$. 

The objective is then to estimate $s_{diag}$ and $s_{nb}$. 
Many cost functions and algorithms have been proposed for learning 
hyperparameters in non-linear dynamic systems; see 
\cite{Nelson2000} for a survey of approaches. 
The typical approach involves iteratively optimizing a cost function, such as 
the prediction error or likelihood, over the hyperparameters and states. 
Wan and Nelson \cite{Wan1996} present a dual EKF approach for optimizing 
prediction error separately over the states and hyperparameters for 
non-linear state evolution models. 
Ghahramani and Roweis \cite{Ghahramani1998} 
use an expectation-maximization (EM) algorithm 
to maximize the likelihood of the observation sequence $W^{(t)}$. 
For the experiments in this paper, 
we assume that the state transition matrix $F^t = I$ 
and choose hyperparameters to minimize 
the mean-squared prediction error.

\subsection{Approximation accuracy}
\label{sec:Approx}
As we noted in Section \ref{sec:Inference_known}, the proposed inference 
procedure makes use of two approximations. 
The first approximation involves modeling 
the block densities $\vec{y}^t$ by Gaussian random variables.
It is simply the Gaussian approximation to the binomial 
distribution and is valid provided that the blocks are sufficiently large 
and dense. 
A rule of thumb often presented in introductory statistics textbooks is 
that the Gaussian approximation to a binomial$(n,p)$ distribution is a 
reasonable approximation provided $np(1-p)>5$. 
In the context of the proposed dynamic SBM, the rule of thumb would correspond 
to $n_{ab}^t \theta_{ab}^t (1-\theta_{ab}^t) > 5$ for all $a,b,t$. 
Recall from \eqref{eq:N_def} that $n_{ab}^t$ varies as the product of the 
number of nodes in classes $a$ and $b$. 
Thus the Gaussian approximation is reasonable even for small 
values of $\theta_{ab}^t$; for example, $16$ nodes in both classes $a$ and 
$b$ is sufficient for $\theta_{ab}^t = 0.02$.

The second approximation involves the linearization of the system dynamics 
in the EKF. 
The state estimate from the EKF is only approximately optimal due to the 
linearization, which is a first-order Taylor approximation about the predicted 
state. 
Other approximately optimal filters have been proposed for non-linear systems, 
including the second-order compensated EKF, the unscented Kalman filter, and 
the particle filter. 
These filters are often found to perform better than the EKF for complex 
non-linear models at the cost of higher computational complexity 
\cite{Gustafsson2012}. 
We argue that the EKF is sufficient for the system model posed by 
\eqref{eq:State_evol_model} and \eqref{eq:Obs_model_SBM}. 
The state transition model \eqref{eq:State_evol_model} is 
linear so the only non-linearity is due to the logistic functions 
$h_i, i = 1, \ldots, k^2$ in the observation model. 
The EKF should work well when the bias and variance of the second-order term 
in the Taylor approximation is negligible compared to the observation noise 
variance \cite{Gustafsson2012}, i.e.
\begin{equation*}
	\begin{split}
	&\frac{1}{4}\left[\tr\left(h''_i\left(\hat{\vec{\psi}}^{t|t-1}\right) 
		R^{t|t-1} \right)^T \tr\left(h''_j\left(\hat{\vec{\psi}}^{t|t-1} 
		\right) R^{t|t-1} \right)\right]_{ij} \\
	&+ \frac{1}{2}\left[\tr\left(h''_i\left(\hat{\vec{\psi}}^{t|t-1}\right) 
		R^{t|t-1} \, h''_j\left(\hat{\vec{\psi}}^{t|t-1}\right) 
		R^{t|t-1} \right)\right]_{ij} \ll \Sigma^t,
	\end{split}
\end{equation*}
where $h''_i(\cdot)$ denotes the Hessian matrix of $h_i$, $[x_{ij}]_{ij}$ 
denotes a matrix $X$ with $(i,j)$th entry given by $x_{ij}$, 
and $A \ll B$ denotes that the 
eigenvalues of $A$ are much smaller than those of $B$. 

\begin{figure}[t]
	\centering
	\includegraphics[width=3.4in]{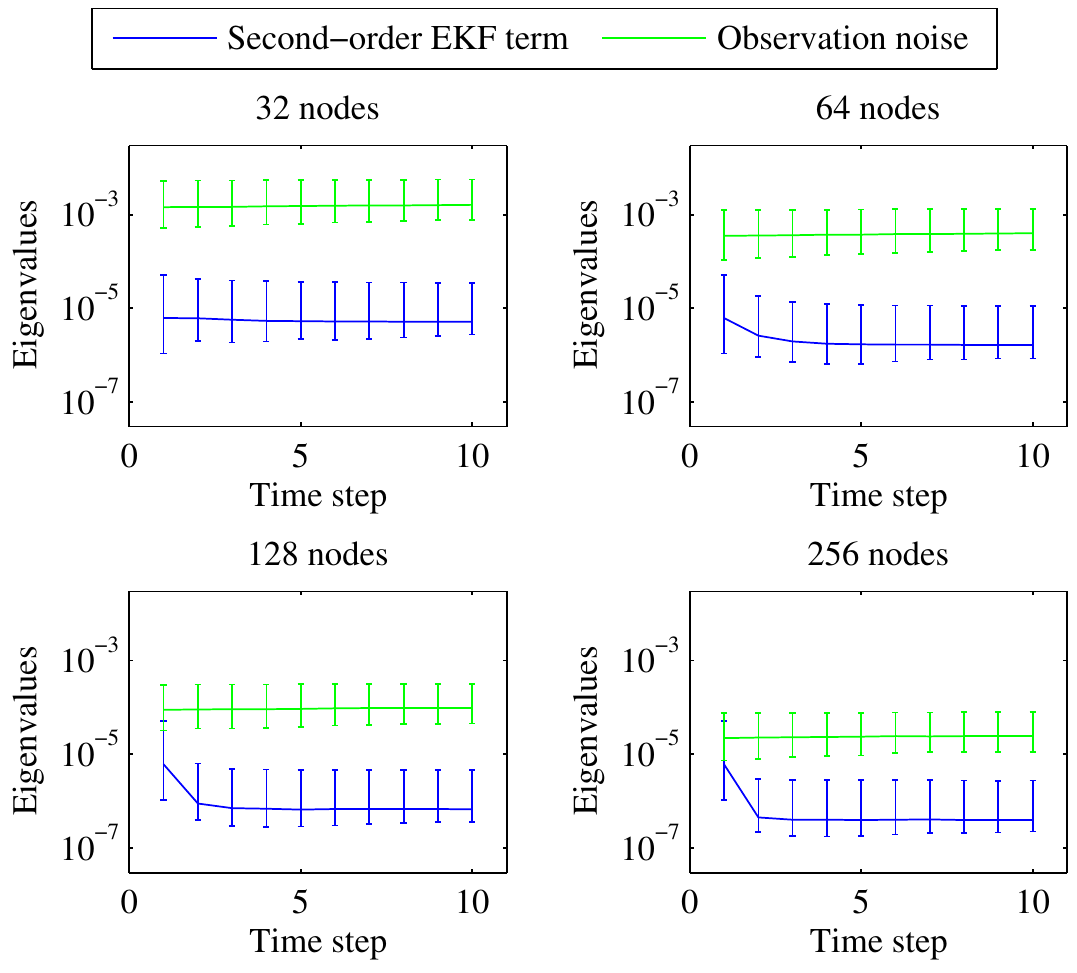}
	\caption{Eigenvalues of second-order EKF term compared to the 
		observation noise variances $(\sigma_{ab}^t)^2$ over time, 
		averaged over $50$ simulation runs. 
		The line denotes the median of the eigenvalues, while the error 
		bars denote the minimum and maximum. 
		The eigenvalues of the second-order EKF term are generally much 
		smaller than the observation noise variances, suggesting that the 
		linear approximation in the EKF should be very accurate.}
	\label{fig:EKF_eigenvalues_n}
\end{figure}

\begin{figure}[t]
	\centering
	\includegraphics[width=3.4in]{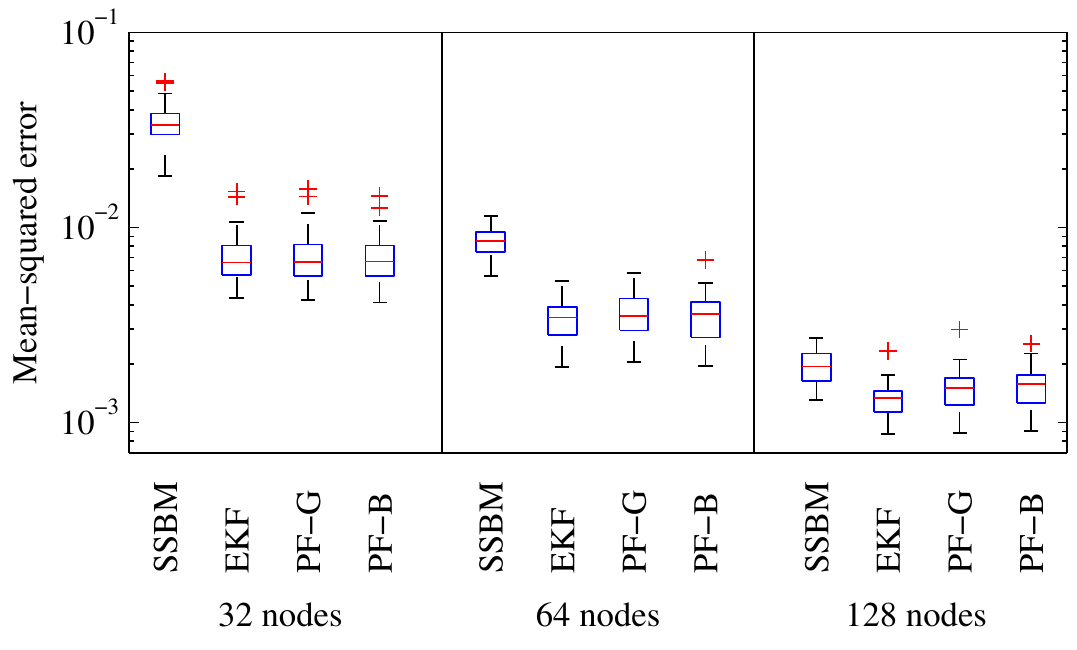}
	\caption{MSE comparison for SSBM, EKF, and two 
		particle filters (PFs) with $10,000$ particles for $50$ simulation 
		runs. 
		PF-G uses the approximate Gaussian observation 
		model, and PF-B uses the actual re-scaled binomial observation model. 
		Edges of the boxes denote $25$th and $75$th percentiles, and the red 
		lines denote medians. 
		Red markers denote outliers.
		The MSE of the EKF is comparable to those of the PFs, confirming the 
		near-optimality of the EKF estimate.}
	\label{fig:SSBM_EKF_PF_comparison}
\end{figure}

We simulate networks drawn from the dynamic SBM to investigate the 
contributions of the second-order EKF term. 
The simulation parameters are chosen based on a synthetic network generator 
proposed by Newman and Girvan \cite{Newman2004}. 
The network consists of four equally sized classes of nodes. 
The mean $\vec{\mu}^0$ of the initial state $\vec{\psi}^0$ is chosen so that 
$\E\left[\theta_{aa}^t\right] = 0.2580$ and $\E\left[\theta_{ab}^t\right] 
= 0.0834$ for $a,b = 1, 2, 3, 4; a \neq b$. 
The initial state covariance $\Gamma^0$ is chosen to be a scaled identity 
matrix $0.04I$. 
$\Theta^t$ evolves according to a Gaussian random walk model on 
$\vec{\psi}^t$, i.e.~$F^t = I$ in \eqref{eq:State_evol_model}. 
$\Gamma^t$ is constructed using \eqref{eq:Gamma_struct}, with 
$s_{diag} = 0.01$ and $s_{nb} = 0.0025$. 
$10$ time steps are generated; at each time, we draw a new graph 
snapshot\footnote{We draw undirected graph snapshots to match the procedure 
in \cite{Newman2004}.} 
from the SBM parameterized by $\Theta^t$ and $\vec{c}^t$. 

Since the observation noise variance \eqref{eq:Sigma_t_Theta} is inversely 
proportional to $n_{ab}^t$, we can simulate a variety of conditions simply by 
varying the number of nodes in the network. 
A comparison of the eigenvalues of the second-order EKF term to the 
observation noise variances for four choices of the number of nodes is shown 
in Fig.~\ref{fig:EKF_eigenvalues_n}. 
We find that the contribution of the second-order term is relatively small, 
suggesting that the first-order Taylor approximation in the EKF is sufficient. 

In Fig.~\ref{fig:SSBM_EKF_PF_comparison}, we present a comparison of the 
mean-squared errors (MSEs) for the static SBM (SSBM), the EKF, and 
two particle filters (PFs), one using the approximate Gaussian distribution for 
$\vec{y}^t$ (PF-G), and one using the actual re-scaled binomial distribution 
(PF-B). 
Each data point in the box plot denotes one simulation run, and MSE refers 
to the mean of the squared tracking error 
$\big\|\hat{\vec{\psi}}^{t|t} - \vec{\psi}^t\big\|_2^2$ over the $10$ 
time steps of the simulation run. 
The MSEs for the EKF and both PFs are comparable, which confirms that the 
EKF is sufficient for the proposed model. 
In addition, there is very little difference in the MSEs for the PF using the 
approximate Gaussian 
distribution and the PF using the actual distribution for $\vec{y}^t$, 
confirming that the Central Limit Theorem approximation is also sufficient. 

\section{Experiments}
\label{sec:Experiments}

\subsection{Simulated stochastic blockmodels}
In this experiment we generate synthetic networks in a manner similar to 
a simulation experiment in \cite{Yang2011}. 
The network consists of $128$ nodes initially split into $4$ classes of $32$ 
nodes each. 
At each time step, $10\%$ of the nodes are randomly selected to 
leave their class and are randomly assigned to one of the other three classes. 
The simulated network parameters are chosen to be the same as in the 
simulated networks described in Section \ref{sec:Approx}. 
$10$ time steps are generated in each simulation run. 

We compare the performance of the EKF-based inference procedure 
to two baselines. 
The first is the static stochastic block model (SSBM) fit to each 
time step individually by spectral clustering \cite{Sussman2012}. 
The second baseline is the probabilistic simulated 
annealing (PSA) algorithm proposed by Yang et al.~\cite{Yang2011}, 
which uses a  
combination of Gibbs sampling and simulated annealing to perform approximate 
inference. 
In the a priori setting, only the EKF and SSBM are applicable, while all 
three methods are applicable in the a posteriori setting. 

\subsubsection{Performance metrics}
\label{sec:SBM_performance}

\begin{figure}[t]
	\centering
	\subfloat[Mean-squared tracking error]
		{\includegraphics[width=1.5in]{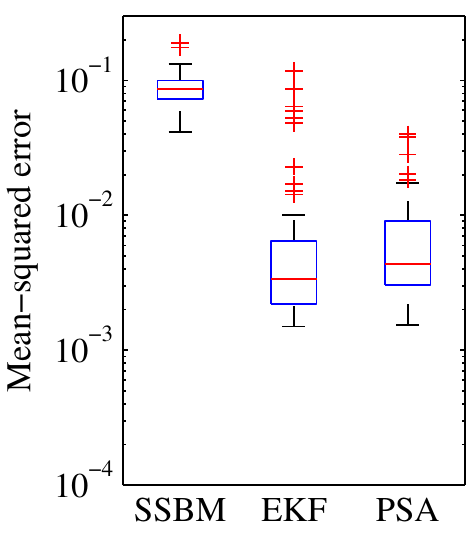}
		\label{fig:SBM_MSE_boxplot}}
	\quad
	\subfloat[Class estimation accuracy]
		{\includegraphics[width=1.5in]{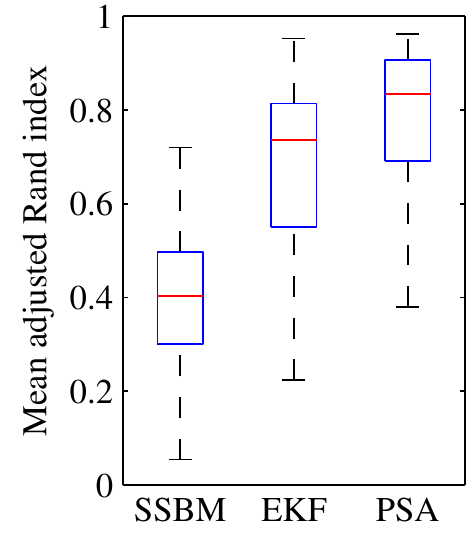}
		\label{fig:SBM_Rand_boxplot}} 
	\caption[Comparison of mean-squared tracking error 
		and class estimation accuracy 
		of a posteriori methods for 
		simulated experiment]
	{Comparison of \subref{fig:SBM_MSE_boxplot} mean-squared tracking error 
		\subref{fig:SBM_Rand_boxplot} and class estimation accuracy 
		of a posteriori methods for simulated experiment. 
	The EKF significantly outperforms the SSBM in both tracking and class 
		estimation.
	The EKF performs slightly better than PSA in tracking and slightly worse 
		in class estimation, but with much less computation time ($45$ seconds 
		for the EKF compared to $365$ seconds for PSA).}
	\label{fig:SBM_boxplots}
\end{figure}

The mean-squared errors (MSEs) for the a priori SSBM and EKF are similar to 
those shown in Fig.~\ref{fig:SSBM_EKF_PF_comparison} for $128$ nodes 
due to the similarity in the experiment setup. 
The MSEs for the a posteriori methods are shown in 
Fig.~\ref{fig:SBM_MSE_boxplot}. 
The proposed EKF method achieves the lowest MSE in both the a priori and 
a posteriori settings.  
The SSBM performs only slightly worse than the EKF in the a priori 
setting since the observation noise variance is inversely proportional to the 
square of the number of nodes in each block. 
However, the SSBM performs extremely poorly in the a posteriori setting due 
to inaccuracy in the estimation of the true classes. 

We evaluate the class estimation accuracy using the adjusted Rand index 
\cite{Hubert1985}.
An adjusted Rand index of $1$ denotes perfect accuracy, and $0$ denotes the 
expected accuracy of a random estimate. 
The adjusted Rand indices for the a posteriori methods are shown in 
Fig.~\ref{fig:SBM_Rand_boxplot}. 
Both EKF and PSA offer a significant improvement over the SSBM approach. 
By approximating the posterior distribution over the classes, the PSA method 
\cite{Yang2011} is able to achieve slightly higher accuracy in estimating the 
true classes compared to our EKF approach, which utilizes a MAP 
estimate of the classes.
However, this comes at the expense of computation time. 
Since the PSA approach uses Gibbs sampling and simulated annealing, 
it requires significantly 
more computation time than both the SSBM and EKF approaches (about  
$6$ minutes for PSA compared to under $1$ minute for EKF and under $1$ second 
for SSBM on a Linux machine with a quad-core 
$3.00$ GHz Intel Xeon processor). 

\subsubsection{Hyperparameter sensitivity}
\label{sec:Hyper_exp}

\begin{figure}[t]
	\centering
	\subfloat[EKF: 
		red asterisk denotes hyperparameter estimates by minimizing 
		mean-squared prediction error.]
		{\includegraphics[width=2.4in]{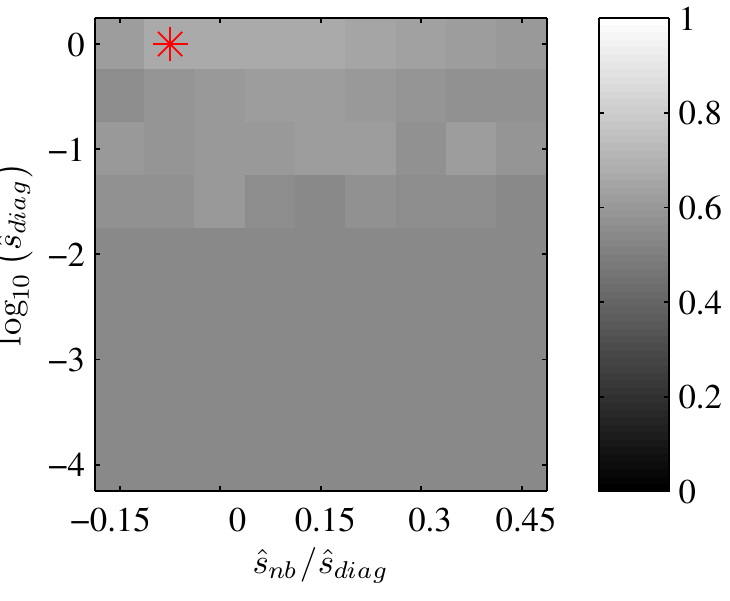}
		\label{fig:SBM_hyper_EKF}} 
	\qquad
	\subfloat[PSA:  
		red asterisk denotes hyperparameter estimates by maximizing 
		modularity. 
		$\hat{\alpha}_{ab}=1$ for all $a \neq b$.]
		{\includegraphics[width=2.4in]{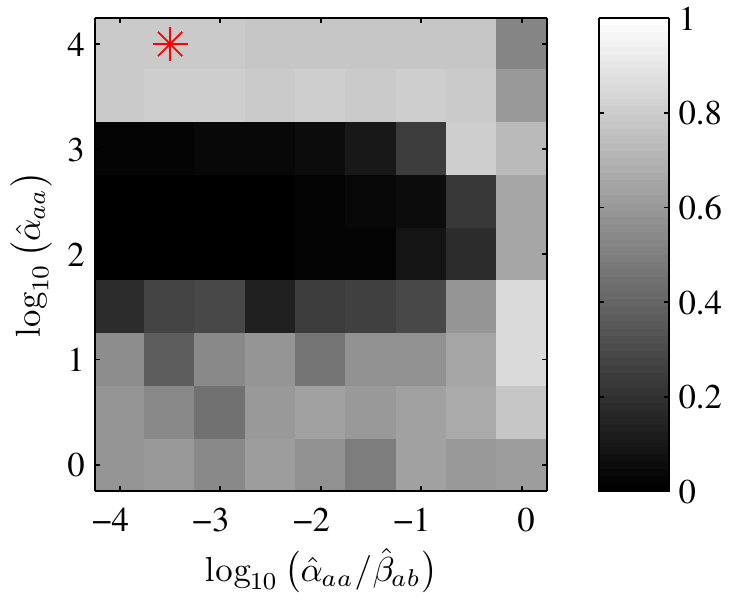}
		\label{fig:SBM_hyper_PSA}}
	\caption{Variation of median adjusted Rand index 
		(over $50$ simulation runs) on 
		hyperparameter settings for EKF and PSA in simulated experiment. 
		The proposed EKF method is robust to the choice of hyperparameters, 
		while the PSA method is extremely sensitive to the choice of 
		hyperparameters.}
	\label{fig:SBM_hyper}
\end{figure}

While PSA is able to slightly 
outperform our proposed EKF method in class estimation accuracy 
(at the cost of higher computation time), it is also more sensitive to 
the choices of hyperparameters. 
Specifically, Yang et al.~\cite{Yang2011} 
note that the accuracy in estimating the 
true classes is sensitive to the choices of the hyperparameters 
for the conjugate prior of the matrix of edge probabilities $\Theta^t$. 
The conjugate prior for each entry $\theta_{ab}^t$ in $\Theta^t$ is Beta 
distributed with hyperparameters $\alpha_{ab},\beta_{ab}$. 
In Fig.~\ref{fig:SBM_hyper}, we plot the variation of adjusted Rand 
index for different choices of hyperparameters for both the a posteriori EKF 
and PSA\footnote{We choose one value of 
$\hat{\beta}_{ab}$ for all $a,b$ and choose $\hat{\alpha}_{ab}=1$ for all 
$a \neq b$, identical to \cite{Yang2011}.}. 
For the EKF, the hyperparameters are $s_{diag},s_{nb}$ as discussed in 
Section \ref{sec:Hyperparameters}. 
Note from Fig.~\ref{fig:SBM_hyper_EKF} that the EKF is robust to the 
choice of hyperparameters, while from Fig.~\ref{fig:SBM_hyper_PSA} it can 
be seen that PSA is extremely sensitive to the choices of 
$\hat{\alpha}_{aa}, \hat{\beta}_{ab}$. 
In particular, certain choices of 
$\hat{\alpha}_{aa}, \hat{\beta}_{ab}$ result in 
mean adjusted Rand indices close to $0$, i.e.~barely better than the 
expected result from randomly assigning nodes to classes. 
Yang et al.~\cite{Yang2011} recommend choosing values of 
$\hat{\alpha}_{aa},\hat{\beta}_{ab}$ that 
maximize the modularity criterion \cite{Newman2004}, which is a measure of 
the strength of community structure for a given partition (when ground truth 
is not available). 
The modularity-based approach is applicable in this experiment because 
the classes correspond to communities, i.e.~$\Theta^t$ is diagonally dominant, 
but is not applicable in the general case where $\Theta^t$ may not be 
diagonally dominant. 
In addition, hyperparameter values that maximize modularity often have 
extremely poor MSE. 
In the general case, one might apply a diffuse prior by setting 
$\hat{\alpha}_{ab} = \hat{\beta}_{ab} = 1$ for all $a,b$. 
Using this approach, the class estimation 
accuracy suffers significantly, as shown in Fig.~\ref{fig:SBM_hyper_PSA}. 

\subsubsection{Scalability}

\begin{figure}[t]
	\centering
	\includegraphics[width=3.4in]{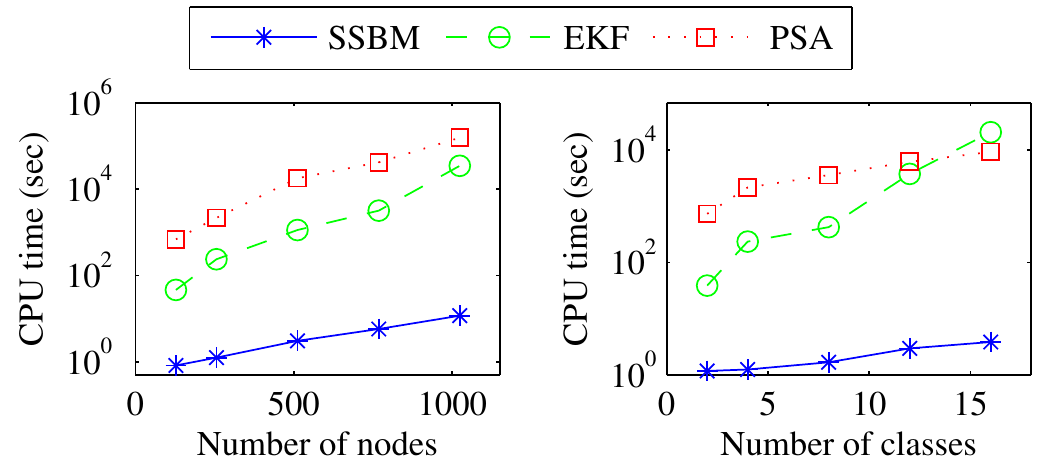}
	\caption{Scalability of a posteriori methods as number of nodes 
		and number of classes is varied in simulated experiment. 
		The EKF method is faster than PSA (except for a large 
		number of classes), which attains 
		similar levels of accuracy, but slower than the SSBM, which has 
		significantly poorer accuracy.}
	\label{fig:SBM_scalability}
\end{figure}

We evaluate the scalability of the EKF-based algorithm and the baselines 
by varying both the number of nodes and the number of classes 
in the experiment. 
Both a priori algorithms require only seconds, 
so we compare only the a posteriori algorithms. 
As noted in Section \ref{sec:Complexity}, the time complexity of the a 
posteriori EKF at each time step is $O(|E^t|+k^6+|V^t|lk^5)$, 
where $l$ denotes the number of local search iterations. 
From \cite{Yang2011}, the time complexity of PSA at each time step is 
$O\big(|V^t|+|E^t|+k^2+l(\mathcal{C}_1|E^t|+\mathcal{C}_2|V^t|)\big)$, 
where $l$ denotes the number of iterations in Gibbs sampling, 
and $\mathcal{C}_1,\mathcal{C}_2$ are constants. 
The left pane of Fig.~\ref{fig:SBM_scalability} shows the variation in 
computation time 
as the number of nodes $|V^t|$ is increased from $128$ to $1,024$ with the 
number of classes held constant at $4$. 
The SSBM is fit using spectral clustering \cite{Sussman2012}, which is 
much faster than the EKF and PSA, but does utilize any temporal model, and 
suffers from poor accuracy in recovering the true states, as we showed in 
Fig.~\ref{fig:SBM_MSE_boxplot}. 
On the other hand, the EKF and PSA are comparable in accuracy but the EKF 
is about \emph{an order of magnitude faster}. 

The right pane of Fig.~\ref{fig:SBM_scalability} shows the variation in 
computation time 
as the number of classes $k$ is increased from $2$ to $16$ with the number 
of nodes held constant at $256$. 
Notice that the EKF is an order of magnitude faster than PSA for $k \leq 8$, 
but slower than PSA once $k$ reaches $16$. 
This is also an expected result because the time complexity of PSA 
is quadratic in $k$, while the EKF requires $k^6$. 
Yang et al.~\cite{Yang2011} model temporal variation in the class memberships 
but not in $\Theta^t$, unlike 
the state-space SBM we use; as a result, PSA has higher 
tracking error than the EKF as shown in Fig.~\ref{fig:SBM_MSE_boxplot}. 
The EKF performs near-optimal state tracking requiring the inversion of a 
full $k^2 \times k^2$ covariance matrix 
as mentioned in Section \ref{sec:Complexity}, which scales poorly in $k$. 
For larger values of $k$, one could achieve significant savings in 
computation time by assuming that the process noise covariance matrix 
$\Gamma^t$ is block-diagonal, which would decouple the dynamics. 
Inference could then be performed by using multiple EKFs with a smaller 
state space. 

\subsection{MIT Reality Mining}
\label{sec:Reality}

\begin{table}[t]
	\centering
	\renewcommand{\arraystretch}{1.2}
	\caption{Comparison of class estimation performance for a posteriori 
		methods applied to the MIT Reality Mining network. 
		The EKF has the highest accuracy and is much faster than PSA.}
	\label{tab:Reality_Rand}
	\begin{tabular}{ccc}
		\hline
		Method & Mean adjusted Rand index & CPU time (sec) \\
		\hline
		SSBM & $0.457$ & $<1$ \\
		EKF & $0.558$ & $15$\\
		PSA & $0.539$ & $244$ \\
		\hline
	\end{tabular}
\end{table}

\begin{figure}[tp]
	\centering
	\subfloat[EKF]
		{\includegraphics[width=3.2in]{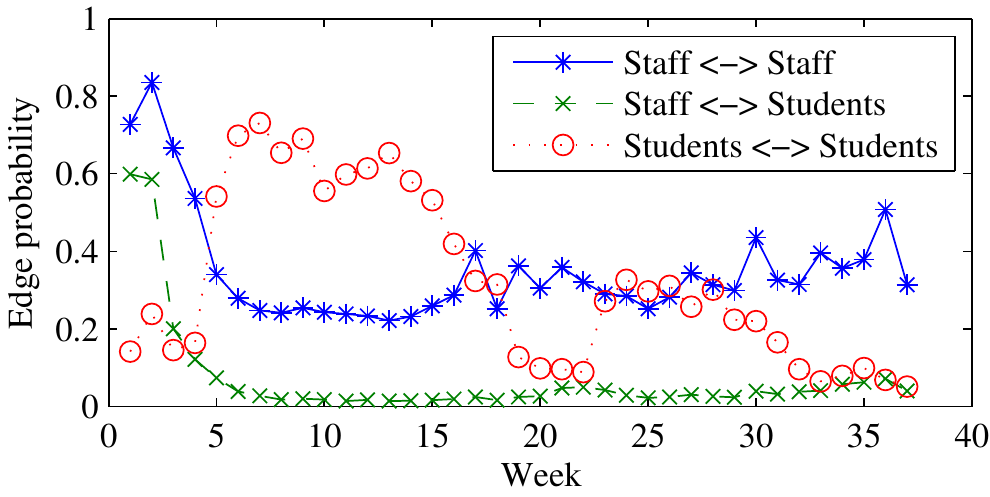}
		\label{fig:Reality_edge_prob_EKF}} \\
	\subfloat[PSA]
		{\includegraphics[width=3.2in]{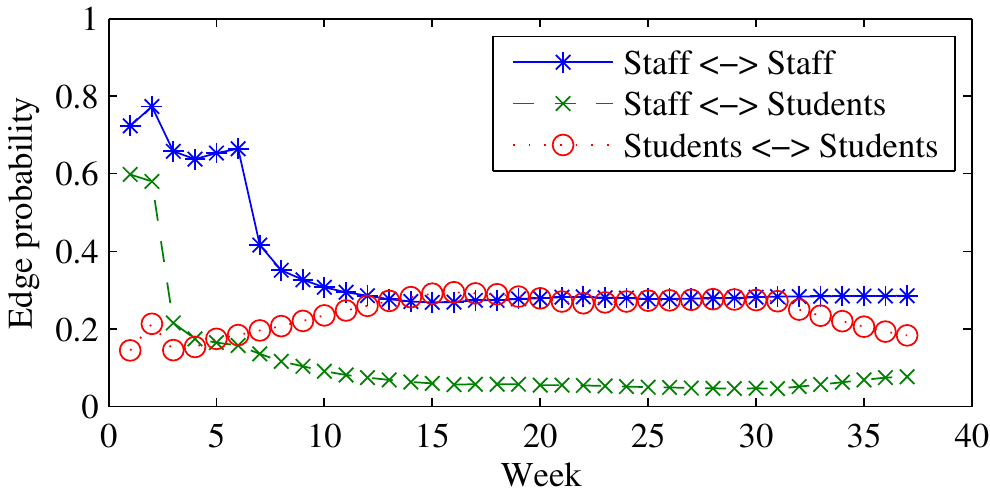}
		\label{fig:Reality_edge_prob_PSA}}
	\caption{Estimated edge probabilities for MIT Reality Mining network. 
		Temporal dynamics corresponding to fall and winter semesters are 
		visible in the EKF estimates of the edge probabilities 
		between students, but not in the PSA estimates.}
		\label{fig:Reality_edge_prob}
\end{figure}

The experiment is conducted on the MIT Reality Mining data set 
\cite{Eagle2009}. 
The data was collected by recording cell phone activity of 94 students and 
staff at MIT over a year. 
We construct dynamic networks based on physical proximity, which was measured 
using scans for nearby Bluetooth devices at $5$-minute intervals. 
We exclude data collected near the beginning and end of the experiment where 
participation was low. 
Each time step corresponds to a $1$-week interval, resulting in $37$ time 
steps between August 2004 and May 2005. 

The affiliation of participants are known, so this data set serves as an 
excellent real data benchmark for dynamic network analysis.
Eagle et al.~\cite{Eagle2009} demonstrated that two communities could be 
found in the time-aggregated network of physical proximity, corresponding to 
first-year business school students and staff working in the same 
building. 
We use these participant affiliations as ground-truth class memberships and 
compare the class estimation accuracy of the a posteriori dynamic SBM 
methods. 
Karrer and Newman \cite{Karrer2011} showed that class 
memberships from a posteriori blockmodeling do not agree with community 
memberships when there is significant degree heterogeneity within communities. 
To reduce the degree heterogeneity, we connect each participant $i$ to the 
$10$ other participants who spent the most time in physical proximity of $i$ 
during each time step.  

A summary of the class estimation performance for the three a posteriori 
methods is shown in Table \ref{tab:Reality_Rand}. 
Both the EKF and PSA, which utilize dynamic models, are more accurate than 
the SSBM fit using spectral clustering. 
Notice that the EKF actually has higher class estimation accuracy than the 
more computationally demanding PSA. 
We find that this is due to the temporal model of the 
edge probability matrix $\Theta^t$ in our proposed dynamic SBM, which PSA 
does not utilize. 
Notice from Fig.~\ref{fig:Reality_edge_prob} that 
PSA does not adapt well to changes in the edge probabilities over time, which 
degrades its class estimation accuracy compared to our proposed EKF method. 

\subsection{Enron email network}
\label{sec:Enron}

We run this experiment on a dynamic social network constructed 
from the Enron corpus \cite{Priebe2005,Priebe2009}, which 
consists of about $500,000$ email messages between $184$ Enron employees 
from 1998 to 2002. 
We place directed edges between employees $i$ and 
$j$ at time $t$ if $i$ sends at least one email to $j$ during week $t$. 
Each time step corresponds to a $1$-week interval. 
We make no distinction between emails sent ``to'', ``cc'', or ``bcc''. 
In addition to the email data, the roles of most of the employees within the 
company (e.g.~CEO, president, manager, etc.) are available, 
which we use as classes for a priori blockmodeling. 
Employees with unknown roles are placed in an ``others'' class. 
We remove the first $56$ and last $13$ weeks of data, where only a few 
emails were sent. 

\subsubsection{Dynamic link prediction}

Unlike in the simulated data set, we do not know the ground truth states 
in this experiment. 
Thus we turn to the task of dynamic link prediction \cite{Tylenda2009} 
to provide a basis for comparison. 
Dynamic link prediction differs from static link 
prediction \cite{Liben-Nowell2007} because the link predictor must 
simultaneously predict the new edges that will be added at time $t+1$, as 
well as the current edges (as of time $t$) that will be removed at time 
$t+1$, from the observations $W^{(t)}$. 
The latter task is not addressed by most static link prediction methods in the 
literature. 

Since the SBM assumes stochastic equivalence between nodes in the same class, 
the EKF and PSA methods alone are only good predictors of the block densities 
$Y^t$, not the edges themselves. 
However, the EKF and PSA methods can be combined with a predictor that 
operates on individual edges to form a good link predictor. 
A simple individual-level predictor is the exponentially-weighted 
moving average (EWMA) \cite{Cortes2003,Xu2011b} 
given by $\hat{W}^{t+1} = \lambda \hat{W}^{t} + 
(1-\lambda) W^t$. 
Using a convex combination of the EKF or PSA and EWMA predictors, we obtain 
a better link predictor that incorporates both 
block-level characteristics (through the EKF or PSA) and individual-level 
characteristics (through the EWMA). 
We evaluate the performance of the link predictors using the area under the 
receiver operating characteristic curve (AUC) metric. 
Since the PSA implementation we used accepts only undirected graphs, we 
reciprocate all edges to create undirected graphs 
for the link prediction experiment. 

\begin{table}[t]
	\centering
	\renewcommand{\arraystretch}{1.2}
	\caption{Comparison of dynamic link prediction performance for the 
		Enron network. 
		The EKF + EWMA and PSA + EWMA approaches perform comparably and 
		better than the EWMA alone.}
	\label{tab:Link_prediction}
	\begin{tabular}{ccc}
		\hline
		Method & AUC & CPU time (sec) \\
		\hline
		EWMA & $0.913$ & $<1$ \\
		A priori EKF + EWMA & $0.939$ & $<1$\\
		A posteriori EKF + EWMA & $0.941$ & $148$ \\
		PSA + EWMA & $0.942$ & $1,440$ \\
		\hline
	\end{tabular}
\end{table}

As shown in Table \ref{tab:Link_prediction}, the dynamic SBM approaches 
combined with the EWMA 
all perform roughly comparably in terms of AUC, and all better than the 
EWMA alone. 
Notice that the a priori EKF adds hardly any computation time to the EWMA. 
If class memberships are not known in advance, one would have to use a 
posteriori methods. 
Both a posteriori methods perform roughly equally in terms of 
AUC, but our proposed EKF method is once again an order of magnitude 
faster than PSA.

\subsubsection{Temporal dynamics}
Next we investigate the temporal variation of the state estimates from the 
EKF. 
Recall that the states $\Psi^t$ correspond to the logit of the edge 
probabilities $\Theta^t$. 
We apply the a priori EKF to obtain the state estimates 
$\hat{\vec{\psi}}^{t|t}$ and their variances (the diagonal of $R^{t|t}$). 
Applying the logistic function, 
we can then obtain the estimated edge probabilities $\hat{\Theta}^{t|t}$ 
with confidence intervals. 

\begin{figure}[tp]
	\centering
	\includegraphics[width=3.45in]{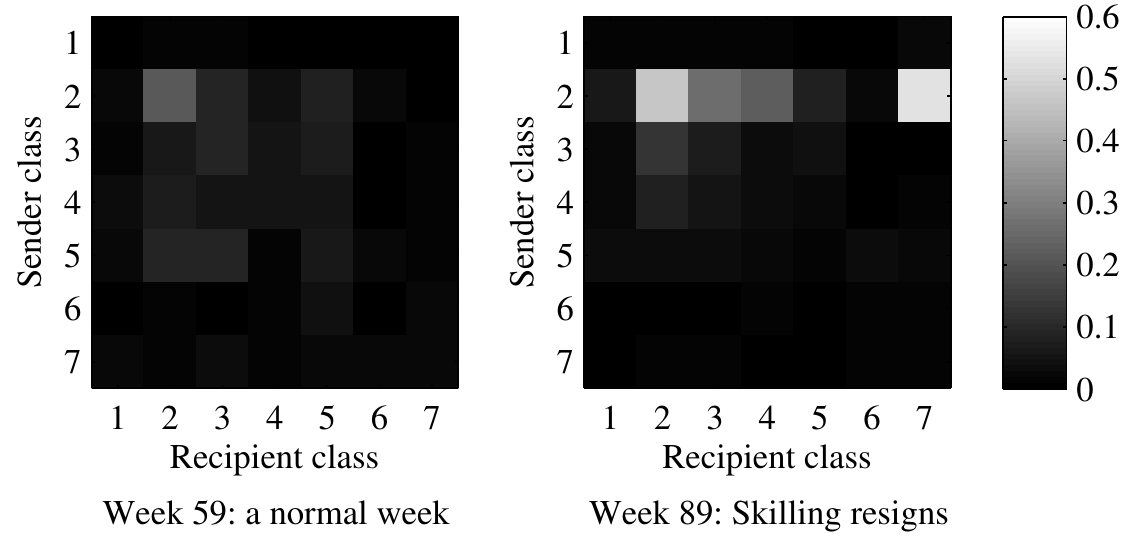}
	\caption{Estimated edge probability matrices for two selected weeks 
		of the Enron email network. 
		Entry $(i,j)$ denotes the estimated probability of an edge from 
		class $i$ to class $j$. 
		Classes are as follows: (1) directors, (2) CEOs, (3) presidents, 
		(4) vice-presidents, (5) managers, (6) traders, and (7) others. 
		Notice the increase in the probability of edges from CEOs during the 
		week of Skilling's resignation.}
	\label{fig:Enron_heatmaps}
\end{figure}

Examining the temporal variation of $\hat{\Theta}^{t|t}$ 
reveals some interesting trends. 
For example, a large increase in the probabilities of edges from CEOs is 
found at week $89$, in which CEO Jeffrey Skilling resigned. 
Inspection of the content of the emails sent during week $89$ 
confirms Skilling's resignation to be 
the cause of the increased probabilities. 
Fig.~\ref{fig:Enron_heatmaps} shows a comparison of the matrix 
$\hat{\Theta}^{t|t}$ during a normal week and during the week Skilling 
resigned. 

\begin{figure}[tp]
	\centering
	\subfloat[CEOs to presidents]
		{\includegraphics[width=3.2in]{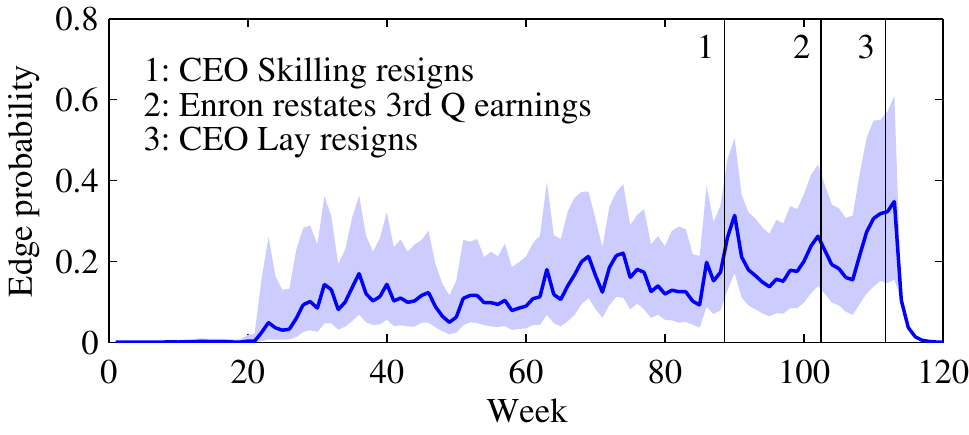}
		\label{fig:Enron_CEOs_pres}} \\
	\subfloat[Others to others]
		{\includegraphics[width=3.2in]{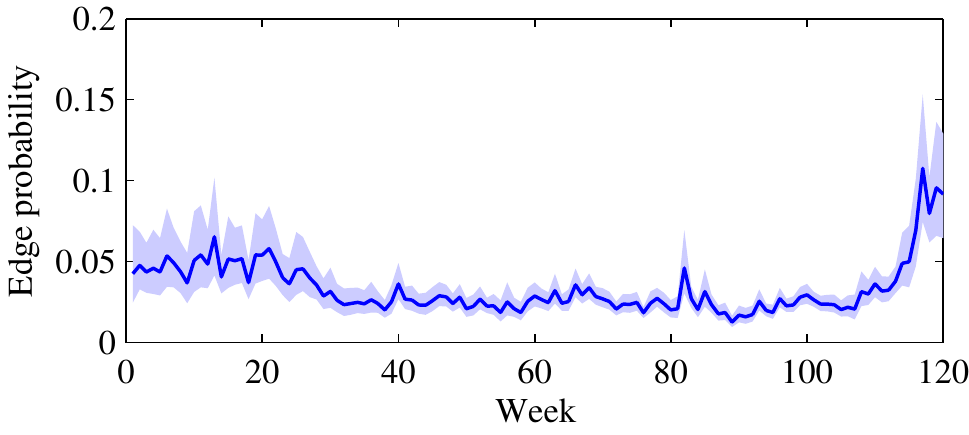}
		\label{fig:Enron_others_others}}
	\caption[A priori EKF estimated edge probabilities]
		{A priori EKF estimated edge probabilities 
		$\hat{\theta}_{ab}^{t|t}$ (solid lines)
		with $95\%$ confidence intervals (shaded region) for selected 
		$a,b$ by week in the Enron network. 
		Estimated edge probabilities 
		from CEOs to presidents peak at times corresponding to 
		three major events (labeled). 
		Edge probabilities between those in 
		other roles increase only after 
		Enron falls under federal investigation.}
		\label{fig:Enron_edge_prob}
\end{figure}

\begin{figure}[t]
	\centering
	\includegraphics[width=3.2in]{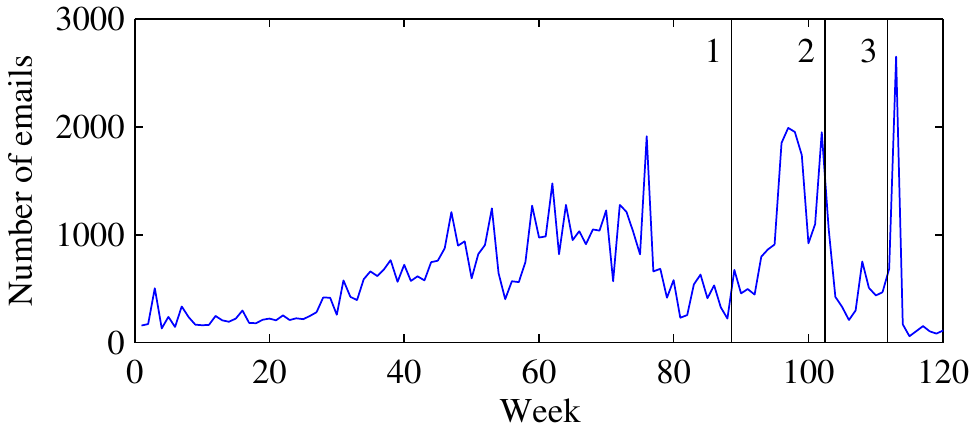}
	\caption{Total number of emails sent each week in the Enron data set. 
		Peaks are found around events 2 and 3 but not event 1.}
	\label{fig:Enron_num_emails}
\end{figure}

Another interesting trend is highlighted in Fig.~\ref{fig:Enron_edge_prob}, 
where the temporal variation of two selected edge probabilities 
over the entire data trace with $95\%$ confidence intervals is shown. 
Probabilities of edges from Enron CEOs to presidents show a 
steady increase as Enron's financial situation worsens, hinting at more 
frequent and widespread insider discussions, while probabilities of edges 
between others 
(not of one of the six known roles) begin to increase only after Enron 
falls under federal investigation. 
Notice also that the estimated edge probabilities from CEOs to presidents 
peak at three times that align with three major events during the 
Enron scandal \cite{PBS2007}. 
A plot of the number of emails sent by week (Fig.~\ref{fig:Enron_num_emails}) 
reveals peaks in email activity around events $2$ and $3$ but not around 
event 1 (CEO Skilling's resignation). 
Specifically, the overall volume of emails did not increase during the week 
Skilling resigned; only the volume of emails originating from CEOs increased, 
as we identified from Fig.~\ref{fig:Enron_heatmaps}. 
Indeed the temporal variation of the edge probabilities 
\emph{between classes}, not the edge probabilities across the entire 
network, is what 
reveals the internal dynamics of this time-evolving social network. 
Furthermore, the temporal model provides estimates with less uncertainty 
than one would obtain by fitting a static SBM to each time step, 
with $95\%$ confidence intervals that are $25\%$ narrower on average.

\section{Conclusion}
\label{sec:Conclusion}

This paper presented a statistical model for time-evolving networks that 
utilizes 
a set of unobserved time-varying states to characterize the dynamics of 
the networks. 
The model extends the well-known stochastic blockmodel for static 
networks to the dynamic setting and can be used for either a priori or a 
posteriori blockmodeling. 
We utilized a near-optimal on-line inference procedure based on the EKF 
that is much faster 
than an existing algorithm based on MCMC sampling yet shows comparable 
accuracy. 

We applied the EKF-based inference procedure to the Enron email network and 
discovered some interesting trends when we examined the estimated states. 
One such trend was a steady increase in edge probabilities 
from Enron CEOs to presidents 
as Enron's financial situation worsened, while edge probabilities 
between other employees 
remained at their baseline levels until Enron fell under federal 
investigation. 
Furthermore, examining the temporal variation of edge probabilities between 
classes revealed a spike in edge probabilities that corresponded to the 
resignation of Enron CEO Jeffrey Skilling; this spike could not be found by 
simply examining the number of emails sent by week.  
The proposed procedure also showed promising results for predicting 
future email activity. 
We believe the proposed model and inference procedure can be applied to 
reveal the internal dynamics of many other time-evolving networks.

\section*{Acknowledgment}
We would like to thank Tianbao Yang for providing the source code 
for the probabilistic simulated annealing algorithm and Kevin Murphy for 
providing the Kalman filtering toolbox for MATLAB. 
We are also grateful to Greg Newstadt for his assistance in the implementation 
of the particle filters. 
Finally we thank Mark Kliger, Brandon Oselio, and the reviewers for their 
comments on the paper.

\bibliographystyle{IEEEtran}
\bibliography{library_sa}

\end{document}